# Affine Defects and Gravitation

By Richard J. Petti
rjpetti@alum.mit.edu

Abstract

We argue that the structure general relativity (GR) as a theory of affine defects is deeper than the standard interpretation as a metric theory of gravitation. Einstein-Cartan theory (EC), with its inhomogenous affine symmetry, should be the standard-bearer for GR-like theories. A discrete affine interpretation of EC (and gauge theory) yields topological definitions of momentum and spin (and Yang-Mills current), and their conservation laws become discrete topological identities. Considerations from quantum theory provide evidence that discrete affine defects are the physical foundation for gravitation.

## 1. Introduction

Einstein Cartan theory (EC) [1,2] is the natural extension of general relativity (GR) to include classical or quantum mechanical intrinsic angular momentum. EC is based on Riemann-Cartan geometry, an extension of Riemannian geometry that includes affine torsion. Riemann-Cartan geometry has as its symmetry group the inhomogeneous rotation group, and it is a continuum theory of translational and rotational defects in affine spaces. These conclusions are elaborated in [1] and [2].

Our objective is to provide terse arguments that (1) EC should be the standard-bearer for GR-like theories; (2) EC is a theory of affine defects whose natural variational principle is Palatini variation with respect to generalized connection coefficients (not with respect to the metric); and (3) physically, EC almost certainly describes *discrete* affine defects.

## 2. General Relativity as a metric theory of gravitation

The standard geometric interpretation of GR is based on the Einstein elevator experiment. The average inward radial acceleration of a bundle of geodesic worldlines is modeled by Ricci curvature, using the Jacobi equation of Riemannian geometry. This leads to the vacuum field equations

(1) $$\mathrm{Ric}_{\alpha\beta} = 0$$

and, accounting for momentum conservation, the field equations of GR

(2) $$E_{\alpha\beta} = 8\pi G\, P_{\alpha\beta}$$

where $\mathrm{Ric}$ = Ricci curvature, $E_{\alpha\beta} = \mathrm{Ric}_{\alpha\beta} - \tfrac{1}{2} (\mathrm{Ric}_{\gamma\delta}\, g^{\gamma\delta})\, g_{\alpha\beta}$ = Einstein tensor, and $P_{\alpha\beta}$ = momentum flux tensor. Riemannian geometry is defined by the metric tensor; the variational principle of the theory varies the metric; and the field equations are solved for the metric. According to this interpretation, GR is a metric theory of gravity.

## 3. Einstein-Cartan theory as an affine theory of gravitation

GR is the master theory of classical mechanics, except it cannot handle spin orbit coupling. Spin occurs in physical theories which are coarse enough that some angular momentum is modeled as intrinsic to a point. Materials scientists know that during spin-orbit coupling the stress 3-tensor becomes non-symmetric, which the Einstein tensor of Riemannian geometry cannot. [1]

In 1922 E. Cartan proposed including non-zero affine torsion in the computations that lead to GR. The result, called Einstein-Cartan theory or EC [1,2] has an algebraic torsion-spin equation, in addition to an equation relating the generalized Einstein curvature to the momentum flux.

(3) $$E_i^\alpha = 8 \pi G P_i^\alpha$$

(4) $$S_{ij}^\alpha = 8 \pi G J_{ij}^\alpha$$

Here, $S = T_{ij}^\alpha + g_i^\alpha T_{j\gamma}^\gamma - g_j^\alpha T_{i\gamma}^\gamma$ = modified torsion tensor, $T_{ij}^\alpha$ = torsion tensor, and $J$ = spin tensor.

In EC theory, it is convenient – and conceptually more correct – to distinguish between base manifold indices (Greek letters) and fiber or current indices (Latin letters). Latin indices are covariant differentiated with respect to the full connection with torsion, and the Greek indices are covariant differentiated with respect to the Levi-Civita connection of the metric g (that is, the torsion-free metric connection of g). In theory, the base space (Greek) indices are not covariant differentiated at all. They always appear in antisymmetric index sets, and they are differentiated only by exterior derivative operators (which are coboundary operators, measuring boundary fluxes per unit volume) which depend only on the differentiable structure and not on any connection. The Levi-Civita connection is merely a computational device to conveniently represent divergences, which are Hodge duals of exterior derivatives, and therefore depend on the metric.[1]

Since 1986, we have known that GR plus classical models of spinning matter – with no additional assumptions – imply affine torsion proportional to spin, and EC or something very close [3].[2] The key insight is to interpret rotational curvature and torsion as affine holonomy per unit area. All known deviations of EC from GR are too small to be measured experimentally. However, since GR plus spin essentially imply EC, henceforth we consider EC, with its enlarged inhomogeneous affine symmetry group, as the standard-bearer for GR.

## 4. Discrete topological models for energy and spin

As presented in detail in [1], Riemann-Cartan geometry is the continuum limit of the theory of affine defects in lattices.

- Torsion models a distribution of "dislocations," in the terminology of materials scientists. Of most interest are screw dislocations, which look like circular parking garage ramps, as represented graphically in Figure 1. The torsion trace $Tr_{ij}^j$ represents a continuum limit of a distribution of edge dislocations, depicted in Figure 2.

---

[1] Reference [1], which is written from the point of view of materials science, does not make these distinctions between types of indices and how they are differentiated (i.e. with exterior derivatives with no connection, and connection-dependent covariant derivatives).

[2] In [3], modified torsion $S_{ij}^\alpha$ is defined in terms of torsion $T_{ij}^\alpha$ by $S_{ij}^\alpha = ½ (T_{ij}^\alpha + g_i^\alpha T_{j\gamma}^\gamma - g_j^\alpha T_{i\gamma}^\gamma)$, whereas the factor of ½ should be omitted. This simple change resolves the factor of 2 discussed in that paper.

- Rotational curvature models a distribution of "disclinations." Disclinations consist of pie-shaped angular wedges which have been excised from the lattice or medium (or added to the lattice if sectional curvature is negative). See Figure 3 for a graphical representation.

We reap several rewards for adopting the hypothesis of discreteness. First among them is that we have topological definitions of momentum, spin, and gravitational radiation.

- Energy-momentum consists of disclinations. In particular, energy in its rest frame consists of excised wedges (angle deficits) in planes with one timelike direction and one spacelike direction.

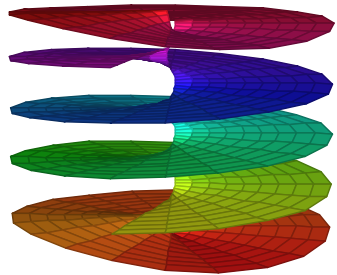

**Fig. 1.** Graphical representation of a screw dislocation

- The interpretation of energy as angular defects extends to the concept of negative energy [4]. Geometrically, negative energy is present at a location where, averaged over all planes with one spatial and one time dimension, there is an excess of angle, instead of a deficit.

- Spin consists of dislocations. For classical rotating black holes and their limiting continuous forms of matter [3] and for Dirac fields [1,2], spin consists of screw dislocations, in the language of materials scientists.

- Weyl curvature, and hence gravitational radiation, consists of distributions of angular wedge deficits and surpluses in different planes, arranged so that no microscopic 3-sphere or 3-pseudosphere has more or less solid angle than is prescribed by the local Ricci curvature.

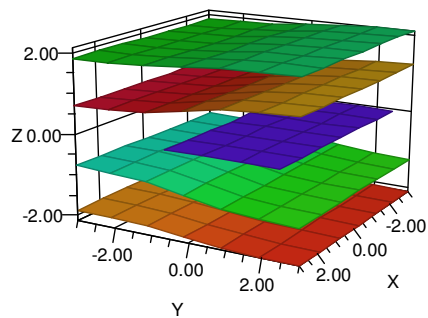

**Fig. 2.** Graphical representation of an edge dislocation

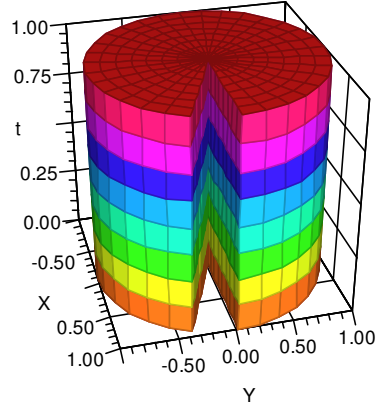

**Fig. 3.** Graphical representation of a disclination

Our second reward is that conservation of momentum and spin become discrete topological identities. Dislocation lines cannot end inside a defected regular lattice. In metallurgy, dislocation lines can only end on grain boundaries (hypersurfaces between nearly perfect single crystals). For example, a screw dislocation cannot end inside a nearly-regular lattice without giving birth to other dislocations at the ends of the screw. Similarly, a disclination line cannot end in a nearly-regular lattice without spawning other disclinations. The (contracted) Bianchi identities of Riemann-Cartan geometry

(5) $$E_i{}^\alpha{}_{;\alpha} - S_{jk}{}^\alpha R^{jk}{}_{\alpha i} + T_{ji}{}^\alpha\ E_\alpha{}^j = 0$$

(6) $$S_{ij}{}^\alpha{}_{;\alpha} + Ric_{[ij]} = 0$$

and more generally for connections

(7) $$D\ F = 0$$

(where $D$ = covariant exterior derivative, $F$ = curvature) are precisely the continuum statements of the law that defect lines cannot end.

## 5. The primacy of affine structure over metric structure

The most elegant derivation of EC is to vary the scalar curvature with respect to the translational frames and rotational connection coefficients, and not by varying the metric. This extends variation of the rotational connection coefficients in GR, which is known as Palatini variation. EC is a theory about affine defects, and not about metric tensors. Since the defect densities – the curvatures – are not free to vary at each point, but must satisfy the conservation laws for defect lines, the fundamental fields are the connection coefficients and not the curvature. Through its inhomogeneous structure group, through its translational and rotational curvature, through its distinction between base space and fiber tensors, EC shows that it is important even at the classical level to treat gravitation as an affine phenomenon.[3]

Corroboration for the primal importance of affine structure (over metric structure) comes from quantum theories of gravity, which seem to require that the preservation of the metric under parallel translation is only a low-energy approximation [5]. At high energies the symmetry group can include conformal transformations which alter volumes, and volume-preserving transformations which alter conformal structure. These types of nonmetricity correspond in affine lattices to inclusions/voids, and shear defects.

---

[3] Reference [1] varies the action with respect to the metric and the contorsion $K_{ijk} = -½ (T_{ijk} + T_{kij} + T_{kji})$. Reference [2] varies with respect to the metric-and-contorsion and also the frame-and-contorsion, then shows that the two sets of variations are equivalent. Since EC is a theory of affine defects, the more meaningful variational principle uses the connection coefficients (frame-and-contorsion) and the resulting stress tensor is the one with deeper conceptual significance.

## 5.1. Extended Regge Calculus

Regge calculus is a discrete formulation of general relativity on N dimensional simplicial manifolds where each 1-simplex is assigned a length; equivalently all N-simplices are compatibly metrized as flat manifolds-with-boundary [6,7].

Regge calculus should be restated in terms of "coordinate bundles" [8] (section 2.3) so it can be extended to gauge theories. Let $\{V_\alpha\}$ be a simplicial decomposition of the base manifold, with each N-simplex carrying a trivial principal bundle with group G.

- A "coordinate bundle" is a family of transition maps $f_{\alpha\beta}: V_\alpha \cap V_\beta \to G$, with $f_{\beta\alpha} = f_{\alpha\beta}^{-1}$, such that when $V_\alpha \cap V_\beta \cap V_\gamma$ is nonempty, $f_{\alpha\beta} f_{\beta\gamma} = f_{\alpha\gamma}$. Coordinate bundles allow gauge transformations (the maps $\lambda_\alpha: V_\alpha \to G$ in [8] section 2.10).
- A connection on such a bundle consists of a collection of maps $\varphi_{\alpha\beta}: V_\alpha \cap V_\beta \to G$, with $\varphi_{\beta\alpha} = \varphi_{\alpha\beta}^{-1}$, but without the compatibility conditions on $V_\alpha \cap V_\beta \cap V_\gamma$. The lack of compatibility of the maps $\varphi_{\alpha\beta}$ around a closed circuit of simplices yields the holonomy around that circuit.

Regge calculus for EC has flat patches glued together by metric-preserving inhomogenous Lorentz transformations $\varphi_{\alpha\beta}$ as above. This extension of Regge calculus yields a direct representation of finite disclination and dislocation defects as rotational and translational holonomy respectively. The core of a defect can occur at any N-2 simplex. The net effect is to supplement the metric information of Regge calculus (which implies interfacial coordinate rotations) with translation information at each interfacial (N-1)-simplex.

A further extension is possible in which the connection is not flat on each N-simplex, yielding a theory like finite element analysis, in which fields vary over each element.

A similar construction with maps $\varphi_{\alpha\beta}: V_\alpha \cap V_\beta \to G$ yields a simplicial model of Yang-Mills theories with group G.

## 5.2. Cellular Automata

The primacy of affine defects in gravitational theory has implications for models of mechanics based on cellular automata [4]. Where such models represent fundamental discrete structures, gravitation should arise from defects in the regular affine structure of the network of cells. It seems too artificial that gravitation would be superimposed as some sort of curvature field – itself a continuum model of defects – on top of a perfectly regular affine lattice.

## 6. Evidence in Quantum Physics for the Hypothesis of Discreteness

The fact that many parts of continuous mathematics have discrete analogues provides scant evidence that the phenomena they are used to model are discrete. However, careful consideration of geometrical aspects of conventional quantum theory strongly suggests that discrete structures underlie mechanics.

- The central fact of quantum mechanics is that continuous geometry breaks down in phase space. In the path integral formulation of quantum mechanics, the uncertainty principle tells us that holonomy comes in discrete units. The holonomy in the uncertainty principle results from parallel translating complex numbers around closed loops in a p-q plane in phase space, for pairs of conjugate variables p and q. In this interpretation of symplectic mechanics, *i* p dq is a connection form on a complex line bundle over phase space, ω := *i* dp ^ dq is its curvature, and dω ≡ 0 is the Bianchi identity.

- In light of EC, spin of Dirac fields consists of screw dislocations whose Burghers vectors have lengths that are multiples of ℏ/2. So the discreteness of spin is another case in which holonomy is discrete.

- String theory and related theories with extended cell structures solve many divergence problems in theories of quantum gravity. However, the discrete structures in most of these theories are still built of differentiable manifolds.

## 7. Beyond affine gauge theories

The continuum limit of distributions of non-affine lattice defects, called dispirations by materials scientists, are non-affine connections, which form the mathematical foundation of Yang-Mills theories. Dispirations are defects in the orientation or structure of crystal basis elements which do not affect the affine or metric geometry of a lattice, as represented graphically in Figure 4. However, interpreting internal gauge symmetries as symmetries of crystal basis elements may not be equivalent to interpreting them as symmetries of compactified dimensions in all cases.

It is sensible to conjecture that all gauge theories are continuum limits of theories of lattice defects. In this view, differentiable structure and real topology are anthropocentric approximations to discrete structures. Differentiable structure consists of an infinitesimal flat affine structure at each point. Real numbers are the topological completion of the rational numbers, which arise from countably infinite subdivisibility of intervals.

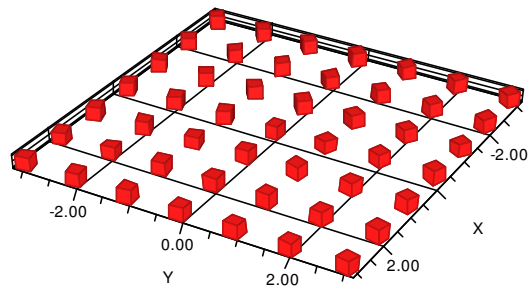

Figure 4: Graphical representation of a dispiration

Concepts of connectedness, parallelism, boundary, and dimension can survive without connections, real topology or differentiable structure. Indeed, all the differential operators in EC and Yang-Mills theories are coboundary operators. EC makes it easier (than GR) to see the coboundary operator structure because it rewards you for distinguishing base space (holonomic, Greek) indices and fiber/current (anholonomic, Latin) indices by becoming computationally much simpler.

An old theme in GR is that all source terms are topological structures in a vacuum. The arguments here suggest that the topological structures of the sources and the fields of EC are dislocations and disclinations, and beyond EC, perhaps dispirations, inclusions, voids, and shear defects. I conjecture that discrete defects will survive EC and gauge theory, because microphysics is discrete, and the deep meaning of curvature in a discrete setting is defect density.